\documentstyle[preprint,prd,aps]{revtex}
\begin{document}

\title{Planar Dirac Electron
in Coulomb and Magnetic Fields}

\author{Choon-Lin Ho$^1$ and V.R. Khalilov$^2$}

\address{\small \sl
1. Department of Physics, Tamkang University, Tamsui 25137, Taiwan\\
2. Department of Physics, Moscow State University\\
Moscow $119899$, Russia}

\maketitle

\begin{abstract}
The Dirac equation for an electron in two spatial dimensions in
the Coulomb and homogeneous magnetic fields is discussed.
This is connected to the problem of the
two-dimensional hydrogen-like atom in the presence of external magnetic field.
For weak magnetic fields, the approximate energy values are
obtained by semiclassical method.
In the case with strong magnetic fields, we present the exact recursion
relations that determine the coefficients
of the series expansion of wave functions, the possible energies and the
magnetic fields.  It is found that analytic solutions are
possible for a denumerably infinite set of magnetic field strengths.
This system thus furnishes an example of the so-called
quasi-exactly solvable models.
A distinctive feature in the Dirac case is that, depending on the strength
of the Coulomb field, not all total angular momentum quantum number
allow exact solutions with wavefunctions in reasonable polynomial forms.
Solutions in the nonrelativistic limit with both attractive and repulsive
Coulomb fields are briefly discussed by means of the method of factorization.
\end{abstract}

\vskip 0.3cm
\noindent{Nov 8, 1999 (3rd Revised version)}
\newpage

\section{Introduction}

Planar nonrelativistic electron systems in a uniform magnetic field are
fundamental quantum systems which have provided insights into many
novel phenomena, such as the  quantum Hall effect and the theory of anyons,
particles obeying fractional statistics  \cite{Prange,Wil}.
Planar electron systems with energy spectrum described by the Dirac
Hamiltonian have also been studied as field-theoretical models for the quantum
Hall effect and anyon theory \cite{NeS}. Related to these field-theoretical
models are the recent interesting studies regarding the instability
of the naive vacuum and spontaneous magnetization in (2+1)-dimensional
quantum electrodynamics, which is induced by a bare Chern-Simons
term \cite{Hoso}.
In view of these developments,  it is essential to have a better understanding
of the properties of planar Dirac particles in the presence of external
electromagnetic fields.

In \cite{KhHo} we studied exact solutions of planar Dirac equation in the
presence of a strong Coulomb field, and the stability of the Dirac vacuum in a
regulated Coulomb field.  Quite recently, there appear interesting studies
on the quantum spectrum of a two-dimensional hydrogen atom in a homogenous
magnetic field \cite{Taut,VP}.  As is well known,
hydrogen atom
in a homogeneous magnetic field has attracted great interest in recent years
because of its classical chaotic behavior and its rich quantum
structures.
The main result
found in \cite{Taut,VP} is that, unlike the three-dimensional case,
the two-dimensional
Schr\"odinger equation \cite{Taut} and the Klein-Gordon equation \cite{VP} can
be solved analytically for a denumerably infinite set of magnetic field
strengths.  The solutions cannot be expressed in terms of special functions
(see also \cite{BGTK}).

In this paper we discuss the motion of Dirac electron in two spatial
dimensions in the Coulomb and homogeneous magnetic fields, and try to obtain
exact solutions of a particular form.
As in the case of the two-dimensional
Schr\"odinger and the Klein-Gordon equation,
by imposing a sufficient condition that guarantees normalizability of the
wavefunctions (see the paragraph after eq.(\ref{r6})), we can obtain the
exact energy levels
for a denumerably infinite set of magnetic fields. In the Dirac case, however,
not all values of the total angular momentum $j$ allow exact solutions with
the form of wavefunctions we assumed here.
Solutions for the nonrelativistic limit of the Dirac equation in 2+1
dimensions are briefly discussed by means of the method of factorization.

We emphasize that in this paper, by assuming an ansatz which guarantees
normalizability of the wavefunction,  only parts of
the energy spectrum of the system are solved exactly.  In particular, we do
not obtain energy levels with magnitude below the mass value, which include
the most interesting ground state solution.  This is the same
as in the Schr\"odinger and the Klein-Gordon case.  All these three cases can
therefore be considered as examples of the newly discovered quasi-exactly
solvable models \cite{TUS}.  In $(3+1)$-dimension,  no analytic solutions, even
for parts of the spectrum, are possible so far.

\section{Motion of Dirac electron in the Coulomb and magnetic fields}

To describe an electron by the Dirac equation in 2+1 dimensions we need only
three anticommuting $\gamma^{\mu}-$matrices.
Hence, the Dirac algebra
\begin{eqnarray}
 \{\gamma^{\mu}, \gamma^{\nu}\} = 2g^{\mu\nu}~,~~~~g^{\mu\nu}= {\rm
diag}(1,-1,-1)
\label{anti}
\end{eqnarray}
may be represented in terms of
the Pauli matrices as $\gamma^0 = \sigma_3$, $\gamma^k = i\sigma_k$, or
equivalently, the matrices $(\alpha_1,\alpha_2)=\gamma^0
(\gamma^1,\gamma^2)=(-\sigma_2,\sigma_1)$ and $\beta=\gamma^0$ \cite{NeS}.
Then the Dirac  equation for an electron minimally coupled to an external
electromagnetic field has the form (we set $c=\hbar=1$)
\begin{eqnarray}
(i\partial_t - H_D)\Psi(t, {\bf r}) = 0,
\label{eq2}
\end{eqnarray}
where
\begin{eqnarray}
  H_D = {\bf \alpha}{\bf P} + \beta m - eA^0 \equiv \sigma_1P_2 -
\sigma_2P_1 + \sigma_3m - eA^0
\label{eq3}
\end{eqnarray}
is the Dirac Hamiltonian, $P_{k} = -i\partial_{\mu} + eA_{\mu}$ is the
operator of generalized momentum of the electron, $A_{\mu}$ the
vector potential of the external electromagnetic field, $m$ the rest mass of
the electron, and $-e~ (e>0)$ is its electric charge.  The Dirac wave function
\begin{eqnarray}
\Psi(t, {\bf r}) =
\left( \begin{array}{c}
\psi_1(t, {\bf r})\\
\psi_2(t, {\bf r})
\end{array}\right)~
\label{eqsp}
\end{eqnarray}
is a two-component function ({\it i.e.} a $2$-spinor). Here $\psi_1(t, {\bf
r})$ and $\psi_2(t, {\bf r})$ are the ``large'' and ``small'' components of
the wave functions.

We shall solve for both positive and negative energy solutions of the Dirac
equation (\ref{eq2}) and (\ref{eq3}) in an external Coulomb field
and a constant homogeneous magnetic field $B>0$ along the $z$ direction:
\begin{eqnarray}
 A^0(r) = Ze/r~ (e>0), \quad A_x = -By/2, \quad A_y = Bx/2~.
\label{e1}
\end{eqnarray}
We assume the wave functions to have the form
\begin{eqnarray}
 \Psi(t,{\bf x}) = \frac{1}{\sqrt{2\pi}}\exp(-iEt)
\psi_l(r, \varphi)~,
\label{e3}
\end{eqnarray}
where  $E$ is the energy of the electron, and
\begin{eqnarray}
\psi_l(r, \varphi) =
\left( \begin{array}{c}
f(r)e^{il\varphi}\\
g(r)e^{i(l+1)\varphi}
\end{array}\right)
\label{eqn6}
\end{eqnarray}
with integral number $l$.
The function $\psi_l(r,\varphi)$ is an eigenfunction of the conserved total
angular momentum $J_z=L_z + S_z = -i\partial/\partial\varphi + \sigma_3/2$
with eigenvalue $j=l+1/2$.  One can of course consider wavefunctions which are
eigenfunctions of $J_z$ with eigenvalues $l-1/2$.  These functions are of the
forms of (\ref{e3}) with $\psi_l$ given by
\begin{eqnarray}
\psi_l(r, \varphi) =
\left( \begin{array}{c}
f(r)e^{i(l-1)\varphi}\\
g(r)e^{il\varphi}
\end{array}\right)~.
\label{eqn7}
\end{eqnarray}
But ansatz (\ref{eqn7}) is equivalent to
ansatz (\ref{eqn6})
if one makes the change $l\to l-1$.  It should be reminded that $l$ is not a
good quantum number.
This is evident from the fact that the two components of $\psi_l$ depend
on the integer $l$ in an asymmetric way.
Only the eigenvalues $j$ of the conserved total angular
momentum $J_z$ are physically meaningful.   For definiteness,  in
the rest of this paper, all statements and
conclusions, whenever angular momentum number $l$ is mentioned, are made with
reference to  ansatz (\ref{e3}) and (\ref{eqn6}).

Substituting (\ref{e3}) and (\ref{eqn6}) in (\ref{eq2}), and
taking into account of the equations
\begin{eqnarray}
 P_x \pm iP_y = -ie^{\pm i\varphi}\left(\frac{\partial}{\partial r} \pm
\left(\frac{i}{r}\frac{\partial}{\partial\varphi}-\frac{eBr}{2}\right)
\right)~,
\label{impul}
\end{eqnarray}
we obtain
\begin{eqnarray}
\frac{df}{dr} - \left(\frac{l}{r} + \frac{eBr}{2}\right)f + \left(E + m +
\frac{Z\alpha}{r}\right)g = 0~,  \nonumber \\
\frac{dg}{dr} + \left(\frac{1+l}{r} + \frac{eBr}{2}\right)g - \left(E - m +
\frac{Z\alpha}{r}\right)f = 0~, \label{sys}
\end{eqnarray}
where $\alpha\equiv e^2=1/137$ is the fine structure constant.
If we let
\begin{eqnarray}
F(r)=\sqrt{r}~f(r)~,~~G(r)=\sqrt{r}~g(r)~,
\end{eqnarray}
eq.~(\ref{sys}) becomes:
\begin{eqnarray}
\frac{dF}{dr} - \left(\frac{l+\frac{1}{2}}{r} + \frac{eBr}{2}\right)F +
\left(E + m + \frac{Z\alpha}{r}\right)G = 0~,  \label{F} \\
\frac{dG}{dr} + \left(\frac{l+\frac{1}{2}}{r} + \frac{eBr}{2}\right)G - \left(
E - m + \frac{Z\alpha}{r}\right)F = 0~. \label{G}
\end{eqnarray}
By eliminating $G$ in (\ref{F}) and $F$ in (\ref{G}), one can obtain the
decoupled second order differential equations for $F$ and $G$.  At large
distances, these equations have the asymptotic forms (neglecting $r^{-2}$
terms):
\begin{eqnarray}
\frac{d^2F}{dr^2} &+& \left[E^2-m^2 - eB(l+1)+\frac{2EZ\alpha}{r}
-\frac{1}{4}(eBr)^2 \right]F=0~,\\
\frac{d^2G}{dr^2} &+& \left[E^2-m^2 - eBl+\frac{2EZ\alpha}{r}
-\frac{1}{4}(eBr)^2 \right]G=0~.
\end{eqnarray}
The last term in these two equations, which is proportional to  $r^2$, may be
viewed as the ``effective confining potential".

The exact solutions and the energy eigenvalues with $0<E< m$
corresponding to stationary states of the Dirac equation (\ref{sys}) with
$B=0$ were found in \cite{KhHo}.
The electron energy spectrum in the Coulomb field has
the form
\begin{eqnarray}
 E = m\left[1 + \frac{(Z\alpha)^2}{(n_r + \sqrt{(l+1/2)^2-(Z\alpha)^2})^2}
\right]^{-1/2}~,
\label{spectrum}
\end{eqnarray}
where the values of the quantum number $n_r$
are : $n_r =0, 1, 2,\ldots$, if $l\ge 0$, and $n_r = 1, 2, 3,\ldots$
if $l<0$.
It is seen that
\begin{eqnarray}
 E_0 = m \sqrt{1-(2Z\alpha)^2}~
\label{ground}
\end{eqnarray}
for $l = n_r = 0$, and  $E_0$ becomes zero at $Z\alpha=1/2$, whereas
in three spatial dimensions $E_0$ equals zero at
$Z\alpha=1$.
Thus, in two spatial dimensions the expression for
the electron ground state energy in the Coulomb field of a point-charge
$Ze$ no longer has a physical  meaning at $Z\alpha=1/2$. It is worth noting
that the corresponding solution of the Dirac equation oscillates near the
point $r\to 0$.

For weak magnetic field the wave functions
and energy levels with $E<m$ can be found from (\ref{F}) and (\ref{G})
in the semiclassical approximation.
We look for solutions of this system in the standard form
\begin{eqnarray}
 F(r) =  A(r) \exp(iS(r))~, \quad  G(r) = B(r) \exp(iS(r))~.
\label{sysqua}
\end{eqnarray}
Here $A(r)$ and $B(r)$ are slowly varying functions. Substituting
(\ref{sysqua}) into (\ref{F}) and (\ref{G}), we arrive at an ordinary
differential equation  for $S(r)$ in the form
\begin{eqnarray}
 \left(\frac{dS}{dr}\right)^2 \equiv Q = E^2-m^2-eB(l+1/2) +
\frac{2EZ\alpha}{r} + \frac{(Z\alpha)^2 - (l+1/2)^2}{r^2} -
\frac{(eBr)^2}{4}~.
\label{eqqua}
\end{eqnarray}
The energy levels with $E<m$ are defined by the formula
\begin{eqnarray}
 \int\limits_{r_{min}}^{r_{max}} \sqrt Q dr = \pi\left(-\sqrt{(l+1/2)^2-
(Z\alpha)^2}
+ \frac{EZ\alpha}{\sqrt{|m^2+eB(l+1/2)-E^2|}}\right)~,
\label{eqqua1}
\end{eqnarray}
where $r_{max}$ and $r_{min}$ ($r_{max}>r_{min}$) are roots of equation $Q=0$.
In obtaining (\ref{eqqua1}),  the term $(eBr)^2$ in $Q$ has been dropped.
If we  require the energy spectrum to reduce to (\ref{spectrum})
when $B=0$,  we must equate the right-hand side of (\ref{eqqua1}) to $\pi
n_r$. As a result we obtain (for $l\ne 0$)
\begin{eqnarray}
 E = \left[m+\frac{eB}{2m}\left(l+\frac{1}{2}\right)\right]\left[1 +
\frac{(Z\alpha)^2}{(n_r+\sqrt{(l+1/2)^2-(Z\alpha)^2})^2}
\right]^{-1/2}~.
\label{spectright}
\end{eqnarray}
In the nonrelativistic approximation the energy spectrum takes the form
\begin{eqnarray}
 E_{non} =  - \frac{(Z\alpha)^2m}{2(n_r+|l+1/2|)^2}
+\frac{eB}{2m}\left(l+\frac{1}{2}\right)~.
\label{spectnonb}
\end{eqnarray}
Semiclassical motion of electron in the magnetic and Coulomb fields
can be characterized by means of
the so-called ``magnetic length'' $l_B=\sqrt{1/eB}$
and the Bohr radius $a_{\rm B}=1/Z\alpha m$ of a hydrogen-like atom of
charge $Ze$.
When the magnetic field is weak so that $l_B\gg a_{\rm B}$, or equivalently,
$B\ll B_{cr}\equiv  (Z\alpha)^2 m^2 /e$,
the energy spectrum
is simply the
spectrum of a hydrogen-like atom perturbed by a
weak magnetic field.  We obtain the Zeeman splitting of atomic spectrum
depending linearly upon the magnetic field strength and the ``magnetic quantum
number'' $l+1/2$.

In strong magnetic field the asymptotic solutions of $F(r)$ and $G(r)$ have
the forms $\exp(-ar^2/2)$ with $a=eB/2$ at
large $r$, and $r^\gamma$  with
\begin{eqnarray}
\gamma = \sqrt{(l+1/2)^2 - (Z\alpha)^2}
\label{gamma}
\end{eqnarray}
 for small $r$.
One must have $Z\alpha <1/2$, otherwise the wave function will oscillate as
$r\to 0$ when $l=0$ and $l=-1$.
In this paper we shall look for solutions of $F(r)$ and $G(r)$ which can be
expressed as a product
of the asymptotic solutions (for small and large $r$) and a series in the form
\begin{eqnarray}
 F(r) = r^{\gamma}\exp(-ar^2/2)\sum_{n=0} \alpha_n r^n~,
\label{series}
\end{eqnarray}
\begin{eqnarray}
 G(r) = r^{\gamma}\exp(-ar^2/2)\sum_{n=0} \beta_n r^n~,
\label{series1}
\end{eqnarray}
with $\alpha_0\neq 0~, \beta_0\neq 0$.
Substituting (\ref{series}) and (\ref{series1}) into (\ref{F}) and (\ref{G}),
we obtain
\begin{eqnarray}
\left[\gamma-\left(l+\frac{1}{2}\right)\right]\alpha_0 &+& Z\alpha\beta_0=0~,
\label{r1}\\
\left[\left(\gamma +1\right)
-\left(l+\frac{1}{2}\right)\right]\alpha_1 &+& Z\alpha\beta_1
+\left(E+m\right)\beta_0=0~,\label{r2}\\
\left[\left(n+\gamma\right)
-\left(l+\frac{1}{2}\right)\right]\alpha_n &+& Z\alpha\beta_n
+\left(E+m\right)\beta_{n-1} - 2a\alpha_{n-2}=0~~(n\geq 2)\label{r3}
\end{eqnarray}
from (\ref{F}), and
\begin{eqnarray}
\left(\gamma +l+\frac{1}{2}\right)\beta_0 &-& Z\alpha\alpha_0=0~,
\label{r4}\\
\left(n+\gamma  + l +\frac{1}{2}\right)\beta_n
&-& Z\alpha\alpha_n
-\left(E-m\right)\alpha_{n-1}=0~~(n\geq 1)~\label{r5}
\end{eqnarray}
from (\ref{G}).

Eq.(\ref{r1}) and (\ref{r4}) allow us to express $\beta_0$ in terms of
$\alpha_0$
in two forms:
\begin{eqnarray}
\beta_0&=&\frac{Z\alpha}{\gamma +l+\frac{1}{2}}~\alpha_0~\\
       &=&-\frac{\gamma -l-\frac{1}{2}}{Z\alpha}~\alpha_0~,
\label{b0}
\end{eqnarray}
which are equivalent in view of the fact that
$\gamma= \sqrt{(l+1/2)^2 - (Z\alpha)^2}$ .
Solving (\ref{r2}) and (\ref{r5}) with $n=1$ gives
\begin{eqnarray}
\alpha_1 & =& -\frac{\left(\gamma +l +\frac{1}{2}\right)(E-m)+ \left(\gamma +l
+ \frac{3}{2}\right)(E+m)}{\left(2\gamma
+ 1\right)\left(\gamma+l+\frac{1}{2}\right)}~Z\alpha~\alpha_0~,\label{a1}\\
\beta_1 &=&  \frac{2\left(\gamma +l \right)E-m}{\left(2\gamma +
1\right)}~\alpha_0~.
\label{b1}
\end{eqnarray}
From (\ref{r5}) one sees that
$\beta_n~ (n\geq 1)$ are obtainable from $\alpha_n$ and $\alpha_{n-1}$.  To
determine the recursion relations for the $\alpha_n$, we simply eliminate
$\beta_n$ and $\beta_{n-1}$ in (\ref{r3}) by means of (\ref{r5}).  This leads
to (for $n\ge 2$):
\begin{eqnarray}
\left(n+\gamma+l -\frac{1}{2}\right)\left(n^2 +
2n\gamma\right)\alpha_n~~~~~~~~~~~~~~~~~~\nonumber\\
+Z\alpha\left[\left(n+\gamma+l-\frac{1}{2}\right)(E-m)
+\left(n+\gamma+l + \frac{1}{2}\right)(E+m)
\right]\alpha_{n-1}~\nonumber\\ +\left(n+\gamma+l
+\frac{1}{2}\right)\left[E^2-m^2-2a\left(n
+ \gamma+l-\frac{1}{2}\right)\right] \alpha_{n-2} = 0~. \label{r6}
\end{eqnarray}

Following \cite{Taut}, we impose the sufficient condition that the series
parts of $F(r)$ and $G(r)$ should terminate appropriately in order to
guarantee normalizability of the eigenfunctions.
It follows from (\ref{r6}) that the solution of $F(r)$ becomes a polynomial of
degree $(n-1)$ if the series given by (\ref{r6}) terminates at a certain
$n$ when $\alpha_n = \alpha_{n+1} = 0$, and $\alpha_m=0 ~(m\geq n+2)$ follow
from (\ref{r6}).
Then from (\ref{r5}) we have $\beta_{n+1}=\beta_{n+2}=\ldots=
0$.  Thus in general the polynomial part of the function $G(r)$ is of one
degree
higher than that of $F$.  Now suppose we have calculated $\alpha_n$ in terms
of $\alpha_0$ ($\alpha_0\neq 0$) from (\ref{a1}) and (\ref{r6}) in the form:
\begin{eqnarray}
\alpha_n=K(l,n,E,a,Z)~\alpha_0~.\label{an}
\end{eqnarray}
Then two conditions that ensure $\alpha_n=0$ and $\alpha_{n+1}=0$ are
\begin{eqnarray}
K(l,n,E,a,Z)=0
\label{K}
\end{eqnarray}
and
\begin{eqnarray}
  E^2-m^2 = 2a\left(n+\gamma+l+\frac{1}{2}\right)~~, ~~~n=1,2,\ldots
\label{E}
\end{eqnarray}
Since the right hand side of (\ref{E}) is always non-negative
\footnote{For $l\geq 0$, this is obvious.  For $l\leq -1$, one has
$-1/2 \leq \gamma+l+\frac{1}{2}\leq 0 $, recalling that $Z\alpha <1/2$.},
we  must have $|E|\geq m$ for the energy.
We note here that, similar to the Schr\"odinger and  the Klein-Gordon case,
the  adopted ansatz guarantees the normalizability of the wavefunction,
but does not provide energy levels with magnitudes below $|E|=m$.

For any integer $n$,  eqs.(\ref{K}) and (\ref{E}) give us a certain number
of pairs $(E,a)$ of energy $E$ and the corresponding magnetic field $B$
(or $a$) which would guarantee normalizability of the wave function.
Thus only parts of the whole spectrum of the system are exactly solved.
The system can therefore be considered as an example of the quasi-exactly
solvable models defined in \cite{TUS}.
In principle the possible values of $E$ and $a$ can be obtained by
first expressing the $a$ (or $E$) in (\ref{K}) in terms of $E$ ($a$) according
to (\ref{E}).  This gives an algebraic equation in $E$ ($a$) which can be
solved for real $E$ ($a$).  The corresponding values of $a$
($E$) are then obtained from (\ref{E}).  In practice the task could be tedious.
We shall consider only the simplest cases below, namely, those with $n=1,~2$
and
$3$. In these cases, the solution of the pair ($E,a$) is unique
for fixed $Z$ and $l$.  In general, for $n>3$, there could exist several
pairs of values ($E,a$) ({\it cf}. \cite{Taut,VP}).
Unlike the
non-relativistic case, here negative energy solutions are possible.
As in the case of the (3+1)-dimensional Dirac equation \cite{Dirac}, the
unfilled
negative energy solutions are interpreted as positrons with positive energies.

We mention once again that all the exact solutions presented below,
including the restrictions for the values of $l$ (or more appropriately,
the values of the conserved total quantum number $j=l+1/2$), are
obtained according to the ansatz (\ref{eqn6}), and (\ref{series}) and
(\ref{series1}) with polynomial parts.  Exact solutions for the other parts of
the energy spectrum, if at all possible, would require ansatz of different
forms which are not known yet.

\subsubsection{$n=1$.}

In this case we have $\alpha_0\neq 0$ and
$\alpha_n =0~ (n\geq 1)$.  From (\ref{a1}) one obtains the energies
\begin{equation}
E=-\frac{m}{2(\gamma+l+1)}~.
\end{equation}
Eq.(\ref{E}) with $n=1$ then gives the corresponding values of magnetic fields
$a$.  This results show that, with the ansatz assumed here,  solution with
positive energy cannot be obtained with
$n=1$.  Furthermore, the previously mentioned requirement that $E\leq -m$ can
only be met with $l<0$.

\subsubsection{$n=2$.}

We now consider the next case, in which
$\alpha_0,~\alpha_1\neq 0$, and $\alpha_n=0 ~(n\geq 2)$.  This also implies
$\beta_n\neq 0 ~(n=0,1,2)$ and $\beta_n =0 ~(n\geq 3)$.  From (\ref{E}),
(\ref{r6}) and (\ref{a1}), we must solve the following set of coupled
equations for the possible values of $E$ and $a$:
\begin{eqnarray}
  E^2-m^2 = 2a\left(2+\gamma+l+\frac{1}{2}\right)~,~~~~~~~~~~~~~\label{En2}
\\ Z\alpha\left[(\Gamma + 1)(E-m) + (\Gamma+2)(E+m)\right]\alpha_1
+2a~ \left(\Gamma  + 2\right)\alpha_0~=0~,\label{r6-2}\\
(2\gamma + 1)\Gamma\alpha_1 + Z\alpha\left[\Gamma
(E-m)+(\Gamma+1)(E+m)\right]\alpha_0=0~.~~~~~~~\label{a1-2}
\end{eqnarray}
Here $\Gamma\equiv \gamma+l+1/2$.
From these equations one can check that
$E$ satisfies the quadratic equation
\begin{eqnarray}
\left[(2\Gamma+1)(2\Gamma+3) -
\frac{2\gamma+1}{(Z\alpha)^2}~\Gamma\right]~E^2+4m(\Gamma+1)~E+
m^2~\left[1+\frac{2\gamma+1}{(Z\alpha)^2}~\Gamma\right]~=~0~.
\label{E2}
\end{eqnarray}
This can be solved by the standard formula.  One must be reminded of the
constraint $|E|\geq m$.
For $l\geq 0$, we can obtain analytic solutions with both positive
and negative energies.  But when $l<0$, analytic solutions can only be
obtained for negative energy $E\leq -m$.
Furthermore, it can be checked that $|E|$ is a monotonic decreasing
(increasing) function of $|l|$ ($Z\alpha$) at fixed $Z\alpha$ ($l$).

For $Z\alpha\ll 1/2$, {\it i.e.} for light hydrogen-like
atoms, we can write down approximate expression for energy
near the mass value,
{\it i.e.} $|E|\simeq m$ .
We can obtain from (\ref{E2}) the approximate values of $E$:
\begin{eqnarray}
E_{+}=m\left[1+
\frac{2(Z\alpha)^2}{(2\gamma+1)\Gamma}~\left(\Gamma+1\right)
\left(\Gamma+2\right)\right]~,~~l\geq 0~,
\label{E3+}
\end{eqnarray}
for positive energies, and
\begin{eqnarray}
E_{-}=-m\left[1+
\frac{2(Z\alpha)^2}{(2\gamma+1)}~\left(\Gamma+1\right)
\right]~,~~l\geq 0~~{\rm and} ~l<0~,
\label{E3-}
\end{eqnarray}
for negative energies (in fact, it can be checked from (\ref{E2}) that for
$l<0$, $E$ is always close to $-m$ for any $Z\alpha <1/2$).

When $Z\alpha$ is close to $Z\alpha=1/2$, we have $|E|\gg m$ for $l\geq 0$.
In this case the energy $E$ can be approximated by:
\begin{eqnarray}
E=\pm m\left[1-\left(Z\alpha\right)^2 \frac{(2\Gamma + 1)(2\Gamma + 3)}
{(2\gamma + 1)\Gamma}\right]^{-1/2}~.
\label{E5}
\end{eqnarray}
A consequence following from this formula is that, for each $l\geq 0$,
there is a critical value of $Z$ beyond which polynomial solution with $n=2$
is impossible.  The critical value of $Z$ for each $l$ is found by setting the
expression in the square-root of (\ref{E5}) to zero.  For $l=0$ and $l=1$,
the critical values of $Z$ are $Z\alpha=1/2.936$ and $1/2.316$, respectively.

In the non-relativistic limit (see Sect. III), it is the upper, or the large,
component $f(r)$ of the Dirac wave function that reduces to the
Schr\"odinger wave function.  Hence, in order to compare with the
results considered in \cite{Taut},
it would be appropriate to study the nodal structures of the function $F(r)$
for positive energy solutions in the limit $E\simeq m$.  It is easy to see from
(\ref{r6-2}) or (\ref{a1-2}) that in this limit, $\alpha_0$ and $\alpha_0$
have opposite signs.  Thus $F(r)$ has only one node in this limit, which is the
same as in the Schr\"odinger case.

\subsubsection{$n=3$.}

For the case of $n=3$, exact solution of (\ref{K}) and (\ref{E}) becomes
much more tedious.  Now the values of $E$ and $a$ are solved by the
following coupled equations:
\begin{eqnarray}
  E^2 - m^2 = 2a\left(\Gamma +
3\right)~,~~~~~~~~~~~~~~~~~~~~~~~~~~~~~~~~~~~~~\label{En3}
\\ Z\alpha\left[(\Gamma + 2)(E-m) + (\Gamma+3)(E+m)\right]\alpha_2
+2a~ \left(\Gamma  + 3\right)\alpha_1~=0~,~~\label{r6-3a}\\
4(\gamma +1)(\Gamma + 1)\alpha_2 +
Z\alpha\left[(\Gamma + 1)(E - m)+(\Gamma+2)(E+m)\right]\alpha_1
+4a~ \left(\Gamma  + 2\right)\alpha_0~=0~,\label{r6-3b}\\
(2\gamma +1)\Gamma\alpha_1 + Z\alpha\left[\Gamma
(E-m)+(\Gamma+1)(E+m)\right]\alpha_0=0~.~~~~\label{a1-3}
\end{eqnarray}

In place of (\ref{E2}) we now have a cubic equation
for the energy $E$.  We shall not attempt to solve it here.  It turns out that
the equation satisfied by $E$ can be reduced to quadratic ones without
linear term in $E$ in
the low energy  ($E\approx m$) and the high energy ($E\gg m$) limit, which
correspond to small and large $Z$, respectively.  The results are
\begin{eqnarray}
E_{+}=m\left[1- \frac{2\left(Z\alpha\right)^2(\Gamma +1)(\Gamma + 2)(\Gamma +
3)} {(2\gamma + 1)\Gamma (\Gamma +2) + 2(\gamma+1)(\Gamma +1)^2}\right]^{-1/2}
\label{E6+}
\end{eqnarray}
and
\begin{eqnarray}
E_{-}=-m\left[1- \frac{2\left(Z\alpha\right)^2(\Gamma +1)(\Gamma + 2)(\Gamma +
3)} {(2\gamma + 1)(\Gamma +2)^2 + 2(\gamma+1)(\Gamma + 1) (\Gamma + 3)}
\right]^{-1/2} \label{E6-}
\end{eqnarray}
for $|E|\approx m$, and
\begin{eqnarray}
E=\pm m\left[1- \frac{2\left(Z\alpha\right)^2(\Gamma +1/2)(\Gamma +3/2) (\Gamma
+ 5/2)(\Gamma + 3)}
{(2\gamma + 1) \Gamma(\Gamma+5/2) (\Gamma +2) + 2(\gamma+1)(\Gamma +1/2)
(\Gamma+1) (\Gamma +3)}\right]^{-1/2}
\label{E7}
\end{eqnarray}
for $|E|\gg m$.
The corresponding values of the magnetic field are obtained by
substituting (\ref{E6+}), (\ref{E6-}),  or (\ref{E7}) into (\ref{En3}).
For $l=-1$, eq.(\ref{E7}) is real only for $1/2.65 < Z\alpha< 1/2$.

As in the $n=2$ case, we shall also investigate the nodal
structures of the function $F(r)$
for positive energy solutions in the limit $E\simeq m$.
The zeros of the polynomial part of $F(r)$ is given by
\begin{equation}
r_0=\frac{1}{2}\left[-\left(\frac{\alpha_1}{\alpha_2}\right)\pm
\sqrt{\left(\frac{\alpha_1}{\alpha_2}\right)^2
- 4\left(\frac{\alpha_0}{\alpha_2}\right)}\right]~.\label{root}
\end{equation}
Note that physical solutions of $r_0$, if exist, must be non-negative.
In the limit $E\simeq m$,
eq.(\ref{r6-3a}) and (\ref{a1-3}) give approximately
\begin{eqnarray}
\frac{\alpha_1}{\alpha_2} &=& -\frac{EZ\alpha}{a}~,\label{ratio1}\\
\frac{\alpha_0}{\alpha_2} &=& - \frac{(2\gamma
+1)\Gamma}{2EZ\alpha(\Gamma + 1)}~\frac{\alpha_1}{\alpha_2}~.\label{ratio2}
\end{eqnarray}
We see from (\ref{ratio1}) that $\alpha_1/\alpha_2 <0$ in this limit.

For negative $l<0$, which implies $-1/2\leq\Gamma<0$, we also have
$\alpha_0/\alpha_2 <0$.  Eq.(\ref{root}) then implies that there is
only one positive zero of $F(r)$.  Hence the wave function has only one node
for $l<0$.

When $l\geq 0$, we have $\Gamma >0$, and hence $\alpha_0/\alpha_2 >0$.
It can be checked from (\ref{En3}), (\ref{E6+}), (\ref{ratio1}) and
(\ref{ratio2}) that $(\alpha_1/\alpha_2)^2
> 4(\alpha_0/\alpha_2)$.  Thus $F(r)$ has two positive zeros.
This is also consistent with the results presented in \cite{Taut}
for the Schr\"odinger case (see also the last part of the following Section).

\section{Non-relativistic limit and method of factorization}

The electron in 2+1 dimensions in the nonrelativistic
approximation is described by one-component wave function. This can easily
be shown in full analogy with the (3+1)-dimensional case.
Let us represent $\Psi$  in the form
\begin{eqnarray}
 \Psi = \exp(-imt)
\left( \begin{array}{c}
\psi\\
\chi
\end{array}\right)~.
\label{equten}
\end{eqnarray}
and substitute (\ref{equten}) into (\ref{eq2}).
This results in, to the first order in $1/c$, the following
Schr\"odinger-type equation (instead of the Schr\"odinger-Pauli equation in
3+1 dimensions):
\begin{eqnarray}
i\frac{\partial\psi}{\partial t} =
\left(\frac{P^2_1 + P^2_2}{2m} + \frac{eB}{2m}-\frac{Ze^2}{r}\right)\psi~,
\label{equ11}
\end{eqnarray}
where, as before,  $P_{k} = -i\partial_{\mu} + eA_{\mu}$ denote the
generalized momentum operators.  The term $eB/2m$ in (\ref{equ11}) indicates
that the electron has gyromagnetic factor $g=2$ as in the $(3+1)$-dimensional
case \cite{Dirac}.

One can now proceed in the same manner as in the Dirac case to solve for the
possible energies and magnetic fields.  We shall not repeat it here.
More simply, we make use of the fact that eq.(\ref{equ11}) differs from the
Schr\"odinger
equation discussed in \cite{Taut} only by the positive spin correction term
$\omega_L=eB/2m$, which is the Larmor frequency.
We thus conclude that the denumerably
infinite set of magnetic field strengths obtained in \cite{Taut} are still
intact, but
the corresponding values of the possible energies are all shifted by an
amount $\omega_L$, {\sl i.e.}
\begin{eqnarray}
E=\omega_L (n+1+l+|l|)~.
\label{SP}
\end{eqnarray}
Simply put, the quantum number $n$ in \cite{Taut} is changed to $n+1$.

Let us note here that the energies and magnetic fields in this case
may also be found by means of a method closely resembling
the method of factorization in nonrelativistic quantum mechanics.
We shall discuss this method briefly below.
Both the attractive and repulsive Coulomb interactions will be considered,
since planar two electron systems in strong external homogeneous
magnetic field (perpendicular to the plane in which the electrons is located)
are also of considerable interest for the understanding of the fractional
quantum Hall effect.
Let us assume
\begin{eqnarray}
 \psi(t,{\bf x}) = \frac{1}{\sqrt{2\pi}}\exp(-iEt+il\varphi)
r^{|l|}\exp(-ar^2/2)Q(r)~,
\label{eqff1}
\end{eqnarray}
where $Q$ is a polynomial, and $a=eB/2$ as defined before.  Substituting
(\ref{eqff1}) into (\ref{equ11}), we have
\begin{eqnarray}
 \left[\frac{d^2}{dx^2} + \left(\frac{2\gamma}{x} - x\right)\frac{d}{dx} +
\left(\epsilon \pm \frac{b}{x}\right)\right]Q(x) = 0~,
\label{eqrnew}
\end{eqnarray}
Here $x=r/l_B$, $l_B=1/\sqrt{eB}$, $\gamma=|l|+1/2$,
$b=2m|Z|\alpha l_B=|Z|\alpha\sqrt{2m/\omega_L}$, and
$ \epsilon = E/\omega_L -(2+l+|l|)$.
The upper (lower) sign in (\ref{eqrnew})
corresponds to the case of attractive (repulsive) Coulomb interaction.
This will be assumed throughout the rest of the paper.

It is seen that the problem of finding spectrum for (\ref{eqrnew})
is equivalent to determining the eigenvalues of the operator
\begin{eqnarray}
 H = - \frac{d^2}{dx^2} - \left(\frac{2\gamma}{x}-
x\right)\frac{d}{dx} \mp \frac{b}{x}~.
\label{hamss}
\end{eqnarray}
We want to factorize the operator (\ref{hamss})
in the form
\begin{eqnarray}
 H =  a^{+}a + p,
\label{hams1}
\end{eqnarray}
where the quantum numbers $p$ are related to the eigenvalues of
(\ref{eqrnew}) by $p=\epsilon$.
The eigenfunctions of the operator $H$ at $p=0$ must satisfy
the equation
\begin{eqnarray}
 a\psi=0~.
\label{eigenf}
\end{eqnarray}
Suppose polynomial solutions exist for (\ref{eqrnew}), say
$Q=\prod\limits_{k=1}^s
(x-x_k)$, where $x_k$ are the zeros of $Q$, and $s$ is the degree of $Q$.
Then the operator $a$ must have the form
\begin{eqnarray}
 a = \frac{\partial}{\partial x} - \sum_{k=1}^s \frac{1}{x-x_k}~,
\label{oper}
\end{eqnarray}
and the operator $a^+$ has the form
\begin{eqnarray}
 a^+ = - \frac{\partial}{\partial x} - \frac{2\gamma}{x} + x -
\sum_{k=1}^s \frac{1}{x-x_k}~.
\label{conoper}
\end{eqnarray}

Substituting (\ref{oper}) and (\ref{conoper}) into (\ref{hams1}) and then
comparing the result with (\ref{hamss}), we obtain the following
set of equations for the zeros $x_k$ (the so-called Bethe {\sl ansatz}
equations \cite{TUS}):
\begin{eqnarray}
 \frac{2\gamma}{x_k} - x_k - 2\sum\limits_{j\ne k}^s\frac{1}{x_j-x_k} = 0~,
\quad k=1,\ldots,s~~,
\label{param}
\end{eqnarray}
as well as the two relations:
\begin{eqnarray}
 \pm b = 2\gamma\sum\limits_{k=1}^s x_k^{-1}~,\quad s=p~.
\label{relat1}
\end{eqnarray}
Summing all the $s$ equations in (\ref{param}) enables us to rewrite the
first relation in (\ref{relat1}) as
\begin{eqnarray}
 \pm b = \sum\limits_{k=1}^s x_k~.
\label{relat2}
\end{eqnarray}
From these formulas we can find the simplest solutions
as well as the values of energy and magnetic field strength.
The second relation in (\ref{relat1}) gives $E=\omega_L (2+s+l+|l|)$,
which is the same as in (\ref{SP}) noting that $n=s+1$.

For $s=1, 2$ the zeros $x_k$ and the values of the parameter $b$ for which
solutions in terms of polynomial of the corresponding degrees  exist can
easily be found from  (\ref{param}) and (\ref{relat2}) in the form
\begin{eqnarray}
 s=1~,\quad x_1 = \pm \sqrt{2|l|+1}~, \quad b = \sqrt{2|l|+1}~,
\phantom{mmmmmmmm} \nonumber \\
 s=2~,\quad x_1 = (2|l|+1)/x_2~, \quad x_2 = \pm (1+\sqrt{4|l|+3})/\sqrt{2}~,
\quad b = \sqrt{2(4|l|+3)}~.
\label{roots}
\end{eqnarray}
From (\ref{roots}) and the definition of $b$ one has
the corresponding values of magnetic field strengths
\begin{eqnarray}
 \omega_L = 2m \frac{(Z\alpha)^2}{2|l|+1}, \quad s=1~, \nonumber \\
 \omega_L = m \frac{(Z\alpha)^2}{4|l|+3}, \quad s=2~,
\label{frequen}
\end{eqnarray}
as well as  the energies
\begin{eqnarray}
  E_1 = \frac{2m(Z\alpha)^2}{2(2|l|+1)}(3+l+|l|)~, \nonumber \\
  E_2 = \frac{m(Z\alpha)^2}{(4|l|+3)}(4+l+|l|)~.
 \label{spectr}
\end{eqnarray}
The corresponding polynomials are
\begin{eqnarray}
 Q_1= x-x_1 = x \mp b~, \phantom{mmmmmm} \nonumber \\
 Q_2= \prod\limits_{k=1}^2 (x-x_k) = x^2 \mp bx + 2|l|+1~.
\label{poly12}
\end{eqnarray}
The wave functions are described by (\ref{eqff1}).
For $s=1, 2$  for the repulsive Coulomb field the wave functions do
not have nodes (for $|l|=0, 1$), i.e. the states described by them are ground
states, while for the attractive Coulomb field the wave function for $s=1$ has
one node (first excited state) and the wave function for $s=2$ has two nodes
(second excited state).

\section{Conclusions}

In this paper we consider solutions of the Dirac equation in two
spatial dimensions in the Coulomb and homogeneous magnetic fields.
It is shown by using semiclassical approximation that for weak magnetic fields
all discrete energy eigenvalues
are negative levels of a hydrogen-like atom perturbed by the
magnetic field.
For large magnetic fields,  analytic solutions
of the Dirac equation are possible for a denumerably infinite
set of magnetic field strengths, if the two components of the wave function
are assumed to have the forms (\ref{series})
and (\ref{series1}) with terminating polynomial parts.
Such forms  will guarantee normalizability of the wave functions.
  We present the exact recursion relations that
determine the coefficients
of the series expansion for solutions of the Dirac equation, the
possible energies and the magnetic fields.
Exact and/or approximate expressions of the energy are explicitly given for
the three simplest cases.  For low positive energy solutions, we also
investigate the nodal structures
of the large components of the Dirac wave functions, and find that they are
the same as in the Schr\"odinger case.
We emphasize that, by assuming a sufficient condition on the wavefunction that
guarantees normalizability,  only parts of the energy spectrum of this system
are exactly solved for.
In this sense the system can be considered a quasi-exactly solvable
model as defined in \cite{TUS}.
As in the Schr\"odinger and the Klein-Gordon case,
energy levels with magnitude below the mass value, which include
the most interesting ground state solution, cannot be obtained by our ansatz.
For the corresponding case in $(3+1)$-dimension, no analytic solutions, even
for parts of the spectrum, are possible.

\vskip 3cm
\centerline{\bf Acknowledgment}

This work was supported in part by the Republic of China through Grant No.
NSC 89-2112-M-032-004.

\vfil\eject

\end{document}